%% file: main.tex
\documentclass{article}
\pdfoutput=1

\usepackage[final, nonatbib]{neurips_2020}

\usepackage[utf8]{inputenc} 
\usepackage[T1]{fontenc}    
\usepackage{url}            
\usepackage{hyperref}
\usepackage{booktabs}       
\usepackage{amsfonts}       
\usepackage{amsmath}
\usepackage{nicefrac}       
\usepackage{microtype}      
\usepackage{amssymb} \usepackage{mathrsfs}
\usepackage{appendix}

\DeclareMathOperator*{\argmax}{arg\,max}
\usepackage{wrapfig}
\usepackage{svg}
\usepackage{subfigure}

\newcommand{\orig}{m}
\newcommand{\counter}{m^{\prime}}
\newcommand{\esolpred}{s_{\orig}}
\newcommand{\esolpredc}{s_{\counter}}

\newcommand{\modelf}{\varphi}

\include{symbols}

\usepackage[font=small,skip=.5pt]{caption}
\title{Explaining Deep Graph Networks with Molecular Counterfactuals}
\author{Danilo Numeroso\\
University of Pisa \\
\texttt{danilo.numeroso@phd.unipi.it}\\
\And
Davide Bacciu\\
University of Pisa \\
\texttt{bacciu@di.unipi.it}\\
}
\begin{document}

\maketitle

\input{worknew}
\clearpage

\bibliographystyle{unsrt}
\bibliography{ref}
\include{appendix}
\end{document}

%% file: symbols.tex
\usepackage{amsmath}
\usepackage{amssymb}
\usepackage{amsfonts}
\usepackage{mathtools}

\newcommand{\CalA}{\mathcal{A}}

\newcommand{\CalC}{\mathcal{C}}

\newcommand{\CalII}{\mathcal{I}}

\newcommand{\CalK}{\mathcal{K}}
\newcommand{\CalL}{\mathcal{L}}

\newcommand{\CalQ}{\mathcal{Q}}
\newcommand{\CalR}{\mathcal{R}}
\newcommand{\CalS}{\mathcal{S}}

\newcommand{\CalY}{\mathcal{Y}}

\DeclareMathOperator{\sgn}{sgn}

\DeclarePairedDelimiter\norm{\lVert}{\rVert}%

\newcommand{\LOneNorm}[1]{\norm{#1}_1}

%% file: worknew.tex
\begin{abstract}
We present a novel approach to tackle explainability of deep graph networks in the context of molecule property prediction tasks, named MEG (Molecular Explanation Generator). We generate informative counterfactual explanations for a specific prediction under the form of (valid) compounds with high structural similarity and different predicted properties. We discuss preliminary results showing how the model can convey non-ML experts with key insights into the learning model focus in the neighborhood of a molecule. 
\end{abstract}

\section{Introduction}
The prediction of functional and structural properties of molecules by machine learning models for graphs is a research field with long-standing roots \cite{micheli07}. Much of current research on the topic relies on Deep Graph Networks (DGNs) \cite{zhou2018graph,DBLP:journals/nn/BacciuEMP20}, as they provide a flexible and scalable means to learn effective vectorial representations of the molecules. This has resulted in a trail of works targeting increasing levels of effectiveness, breadth and performance in the prediction of chemo-physical properties \cite{pmlr-v70-gilmer17a}. The scarce intelligibility of such models and of the internal representation they develop can, however, act as a show-stopper for their consolidation, e.g. to predict safety-critical molecule properties, especially when considering well known issues of opacity in DGN assessment \cite{Errica2020A}. In this respect, attention is building towards the development of interpretability techniques specifically tailored to DGNs. While some DGN shows potential for interpretability {\it by-design} thanks to its probabilistic formulation
\cite{bacciujmlr2020}, the majority of works in literature take a neural-based approach which requires the use of an {\it external} model explainer.  GNNExplainer \cite{ying2019gnnexplainer} is the front-runner of the model-agnostic methods providing local explanations to neural DGNs in terms of the sub-graph and node features of the input structure which maximally contribute to the prediction. RelEx \cite{zhang2020relex} extends GNNExplainer to surpass the need of accessing the model gradient to learn explanations. GraphLIME \cite{huang2020graphlime} attempts to create locally interpretable models for node-level predictions, with application limited to single network data. This paper fits into this pioneering field of research by taking a novel angle to the problem, targeting the generation of interpretable insights for the primary use of the experts of the molecular domain. We build our approach upon the assumption that a domain expert would be interested in understanding the model prediction for a specific molecule based on differential case-based reasoning against counterfactuals, i.e. similar structures which the model considers radically different with respect to the predicted property. Such counterfactual molecules should allow the expert to understand if the structure-to-function mapping learned by the model is coherent with the consolidated domain knowledge, at least for what pertains a tight neighborhood around the molecule under study.
We tackle the problem of counterfactual molecule generation by introducing an explanatory agent based on reinforcement learning (RL) \cite{sutton1998reinforcement}. This explanatory agent has access to the internal representation of the property-prediction model as well as to its output and uses this information to guide the exploration of the molecular structure space to seek for the nearest counterfactuals. Our approach is specifically thought for molecular applications and the RL agent leverages domain knowledge to constrain the generated explanations to be valid molecules. We test our explainer on DGNs tackling the prediction of toxicity (classification task) and solubility (regression task) of chemical compounds. 

\section{Molecular Explanation Generator (MEG)}\label{sect:model}
\begin{wrapfigure}{r}{0.5\textwidth}
\vspace{-15pt}
    \centering \includegraphics[width=.99\linewidth]{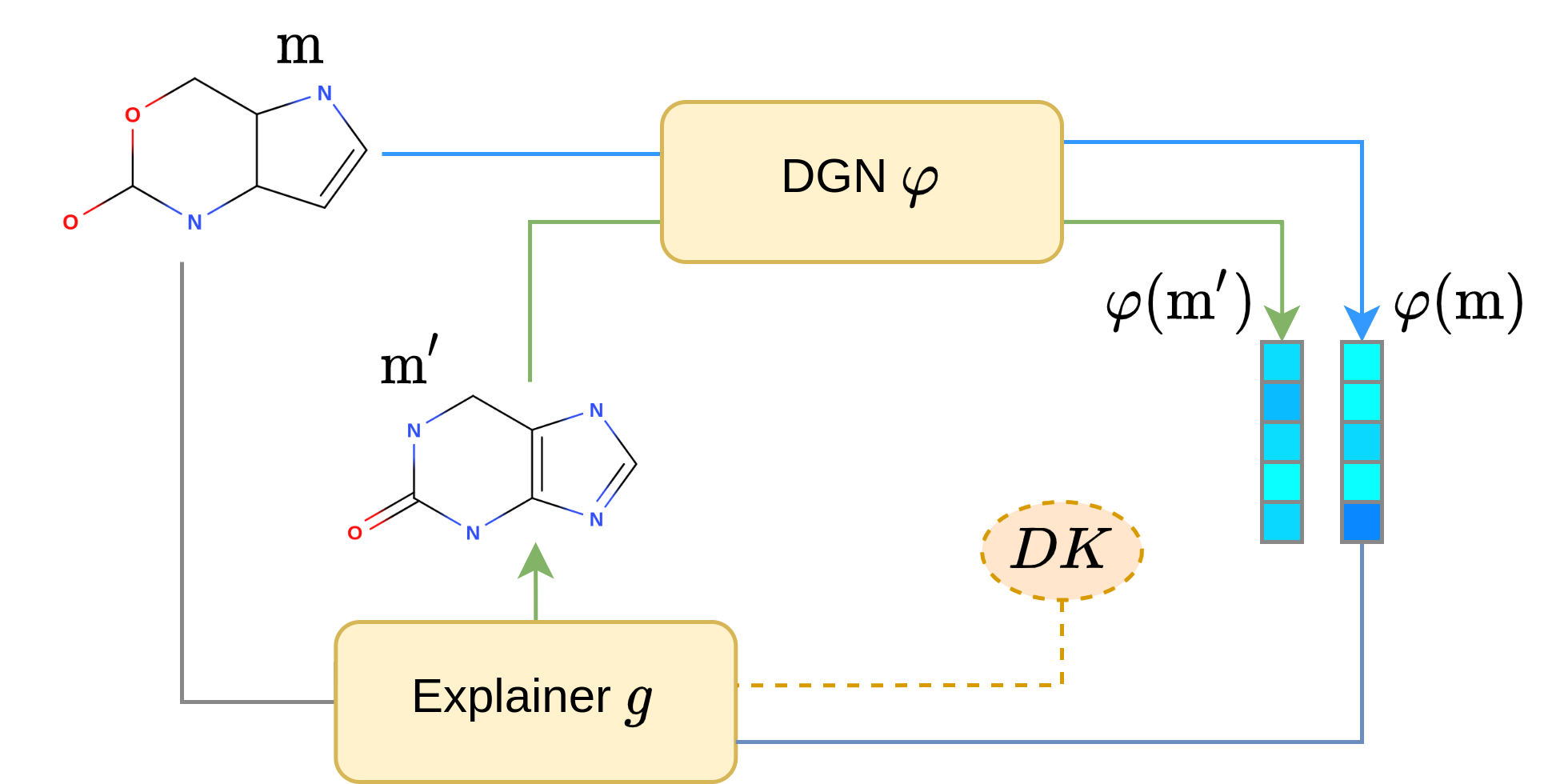}
    \small
    \caption{
        DGN $\varphi$ is a trained molecule property predictor, whereas the Explainer $g$ is a generative agent producing counterfactuals, constrained by prior domain knowledge $DK$.}
    \label{fig:architecture}
\end{wrapfigure}
The overall architecture of our explanation framework, named MEG, is depicted in 
\autoref{fig:architecture}. Here we denote with $\modelf: \CalII \rightarrow \CalY$ a DGN that is fit to solve a molecular property prediction task. $\CalII$ represents the space of (labelled) molecule structures and $\CalY$ is the task-dependent output space. 
The Explainer is an RL agent implementing a generative function
$g: \CalII \rightarrow \CalII$ targeting the generation of counterfactual explanations.  Molecular counterfactuals ought to satisfy three properties: (i) they need to resemble the molecule under study; (ii) predicted properties on counterfactuals must differ substantially from those predicted on the input one; (iii) molecular counterfactuals need to be in compliance with chemical constraints. To this end, the agent $g$ receives information about an input molecule $\orig$ and its associated prediction score $\modelf(\orig)$, and generates a molecular counterfactual $\counter$, leveraging prior domain knowledge to ensure validity of the generated sample. Counterfactual generation is formalised as a maximisation problem in which, given a target molecule $\orig$ with prediction $\modelf(\orig)$, the generator $g$ is trained to optimize:
\begin{equation}\label{eq:L}
    \argmax_\theta \CalL \big( \varphi(\orig), (\varphi \circ g)(\cdot
    \mid \theta) \big) + \CalK \big [\orig, g(\cdot \mid \theta)
      \big].
\end{equation}
The composition $(\varphi \circ g)(\cdot \mid \theta)$ formalizes the model $\varphi$ counter-predictions, made over the counterfactuals produced by $g$. Given the counterfactual $\counter = g(\cdot \mid \theta)$ we rewrite Equation \ref{eq:L} as 
\begin{equation}
\argmax_{m^{'}} \CalL \big( \varphi(\orig), \varphi(\counter) \big) +
\CalK \big [\orig, \counter \big]
    \label{eq:counterfactual-gen}
\end{equation}
where $\CalL$ is a measure of prediction disagreement between the molecule $\orig$ and its counterfactual $\counter$, while $\CalK$ measures $(\orig,\counter)$ similarity. In our framework, $\orig$ is used to bootstrap the generative process in $g$ which operates on the current candidate counterfactual with graph editing operations under domain knowledge constraints. Given the non-differentiable nature of the graph alterations, we model $g$ through a multi-objective RL problem \cite{liu2015morl}, 
that takes the form of an MDP($\CalS$, $\CalA$, $\CalQ$, $\pi$, $\CalR$, $\gamma$). Apart from well known differentiability issues of graph operations, the generator $g$ is modeled as an RL agent for its ease in modelling and handling multi-objective optimization. This allows to easily steer towards the generation of counterfactuals optimizing several properties at a time \cite{sanchez2017,popova2018}. Since we are interested in generating counterfactuals that are compliant to chemical knowledge, the action space $\CalA$ is restricted so as to only retain actions that preserve the chemical validity of the molecule. To this end, we base the implementation of our agent on the MolDQN \cite{zhou2018optimization} model, that is an RL-based approach to molecule graph generation leveraging double Q-learning \cite{hasselt2015deep}. At each step, the reward function $\CalR$ exploits the prediction from $\varphi$ so as to notify the agent of its current performance, emitting a scalar reward.
In our design, $\CalR$ binds together a term regulating the change in prediction scores, which is inherently task-dependent, with a second term controlling similarity between the original molecule and its counterfactual, as presented in Equation \ref{eq:L}.
Currently, we have explored two formulations for the latter term. The former leverages the Tanimoto similarity over the Morgan fingerprints
\cite{rogers2010}. The latter is a $\modelf$-model dependent metric exploiting the encoding of molecules in the DGN internal representation. An advantage of using the latter approach is that it takes into account the model's own perception of structural similarity between molecules.

The leftmost term $\CalL$ in \autoref{eq:counterfactual-gen} can be specialized
for classification and regression tasks. As regards classifications, given a set of classes $\CalC$, a model $\modelf$ emits a probability distribution $\modelf (\cdot) = \boldsymbol{y} = [y_0, ..., y_{|\CalC|}]$ over the predicted classes. In this case, given an input-prediction pair $\langle m,  c = \argmax_{c \in \CalC} \modelf(m) \rangle$, 
the generator is trained to produce counterfactual explanations $\counter$ 
minimising the prediction score for class $c$, as follows
\begin{align}
    \argmax_{\counter} -\alpha y_c
    + (1 - \alpha) \CalK [\counter, \orig]
      \label{eq:tox21-obj1}
\end{align}
where $\alpha \in [0,1]$ is a hyper-parameter weighing the two parts. 
Hence, the model $\modelf$ returns at each step a smooth reward,
which is actually the inverse of the probability of $\counter$
belonging to class $c$. Differently, for a regression task, the objective function can be defined as
\begin{equation}
    \argmax_{\counter} \alpha \sgn\big(\LOneNorm{\esolpredc - s} -
    \LOneNorm{\esolpred - s}\big) \LOneNorm{\esolpredc - \esolpred} +
    (1 - \alpha) \CalK[\counter, \orig] \label{eq:esol-obj}
\end{equation}
where $sgn$ is the sign function, $s$ is the regression target,
and $s_m$ and $s_{\counter}$ are the predicted values for the
original molecule and its counterfactual, respectively.
The sign function is needed to prevent the agent from generating molecules whose predicted scores move towards the original target, by providing negative rewards.

The main use of counterfactual explanations is to provide  insights into the  function learned  by  the  model $\varphi$. In this sense,  a  set of counterfactuals  for a molecule may be used to: (i) identify changes to the molecular structure leading to substantial changes in the properties, enabling domain experts to discriminate whether the model predictions are well founded; (ii) validate existing interpretability approaches, by running them on both the original input graph and its related counterfactual explanations. The main idea behind this latter point is that a local interpretation method may provide explanations that work well within a very narrow range of the input, but do not give a strong suggestion on a wider behaviour. To show evidence and usefulness of such a differential analysis, in the following section we use our counterfactuals to assess the quality of explanations given by GNNExplainer \cite{ying2019gnnexplainer}. Given the undirected nature of the graphs in our molecular application, we restrict the original GNNExplainer model to discard the effect of edge orientation on the explanation.

\section{Experimental Evaluation}
\begin{figure}
    \centering
    \subfigure[A0]{
         \centering
         \includegraphics[width=.21\linewidth]{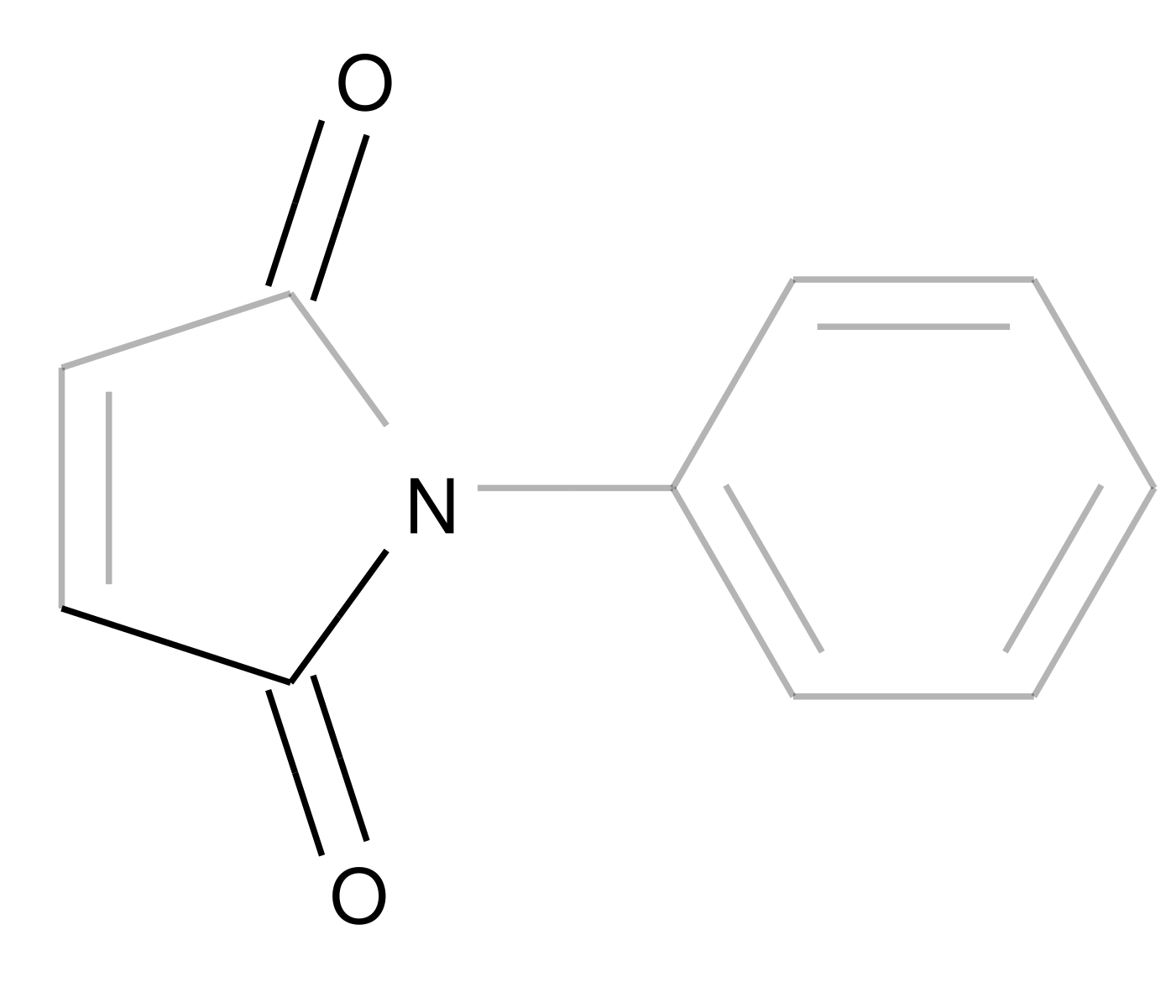}
    }
    \subfigure[A1]{
         \centering
         \includegraphics[width=.21\linewidth]{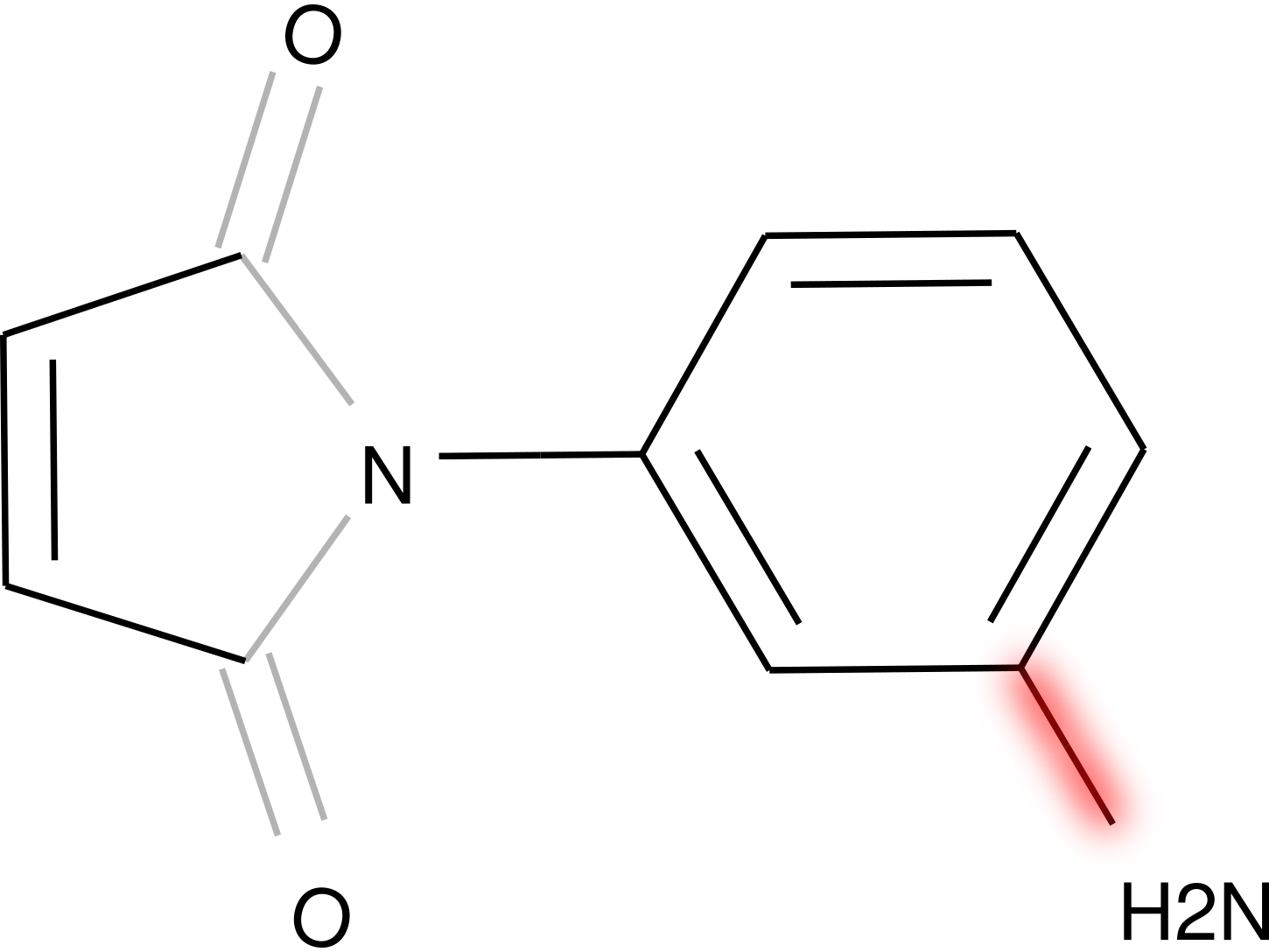}
    }
    \subfigure[A2]{
         \centering
         \includegraphics[width=.19\linewidth]{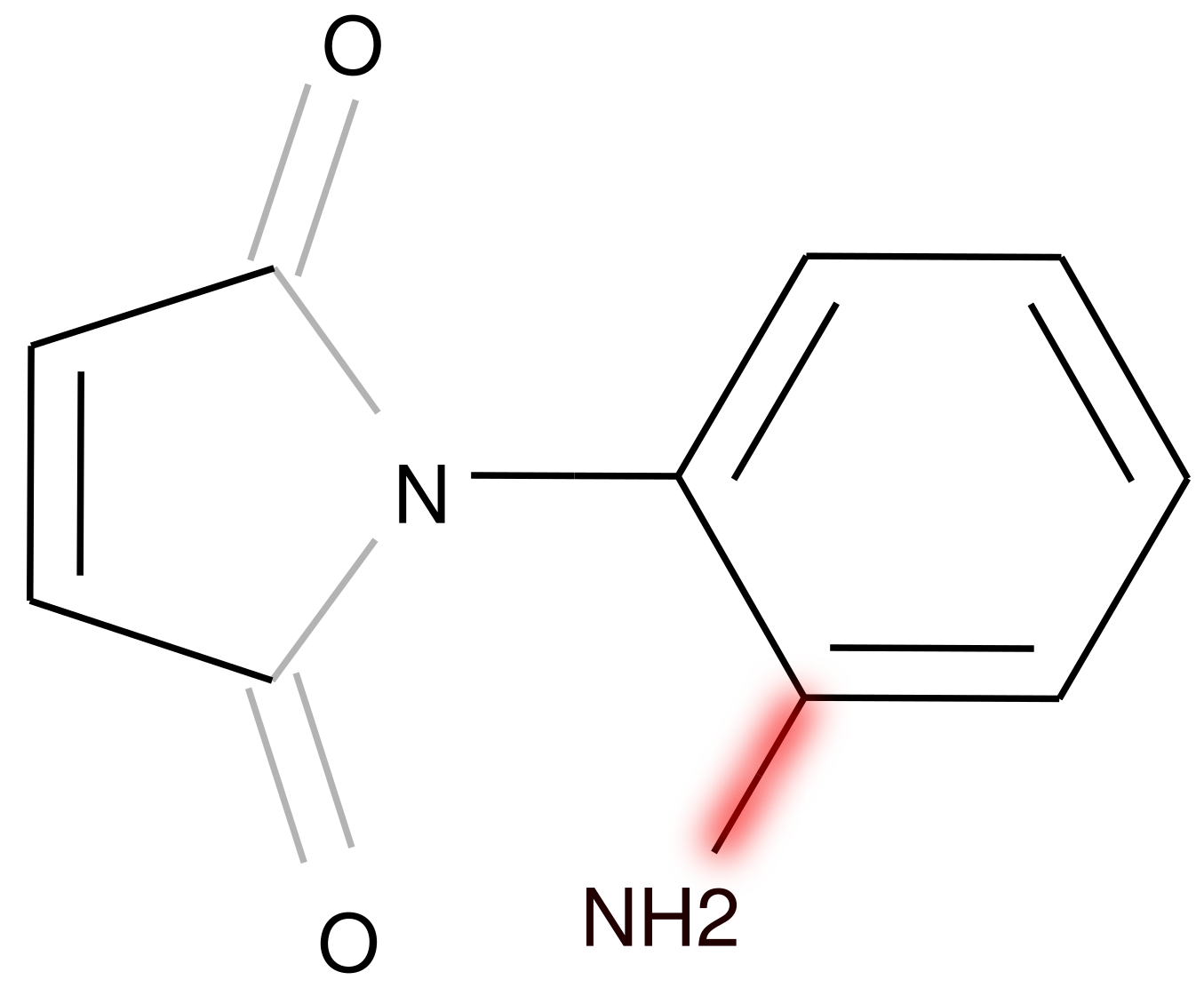}
    }
    \subfigure[A3]{
         \centering
         \includegraphics[width=.19\linewidth]{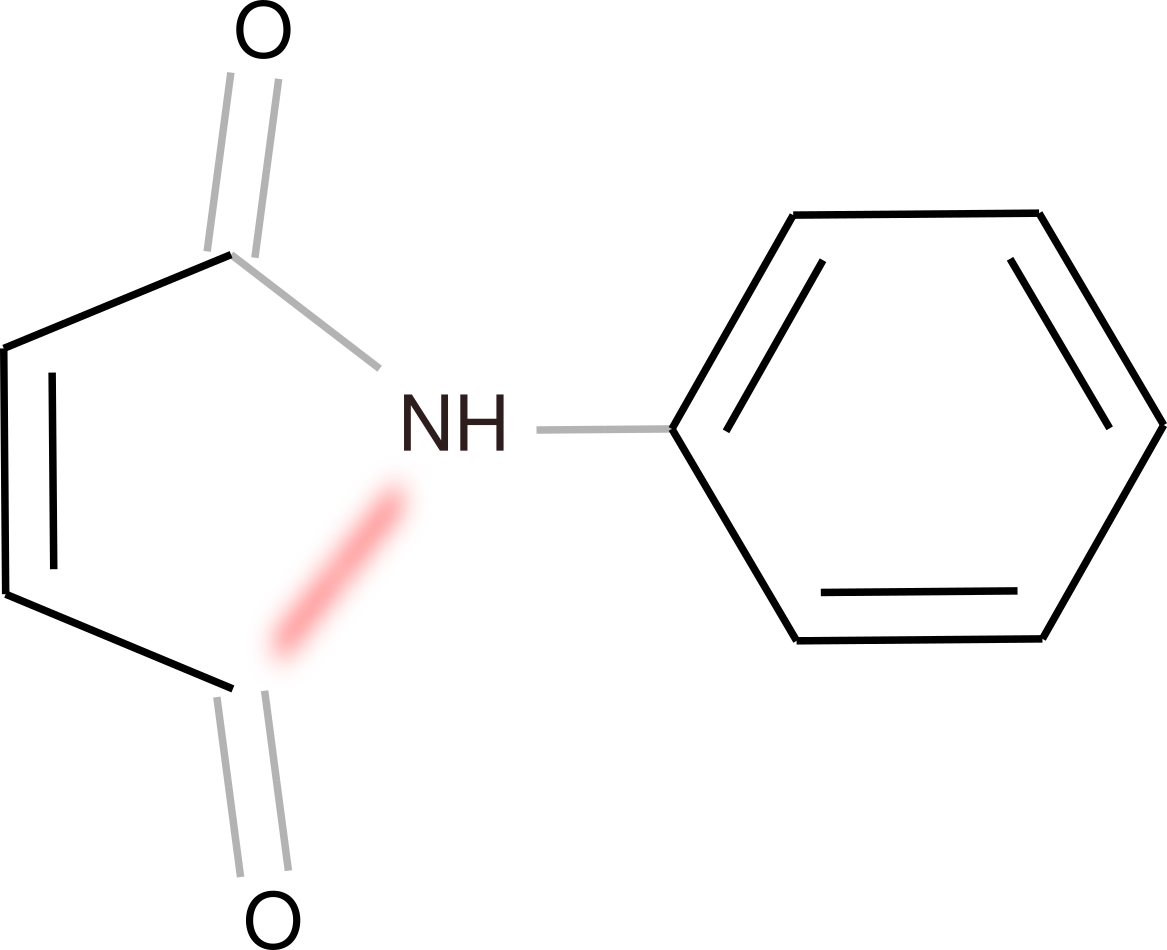}
    }
    \caption{
        Experimental results for the Tox21 sample, reported in \autoref{tab:res}.
    }
    \label{fig:ex1}
\end{figure}
\begin{wrapfigure}{r}{0.5\textwidth}
\vspace{-32pt}
\centering
    \subfigure[B0]{
         \centering
         \includegraphics[width=.35\linewidth]{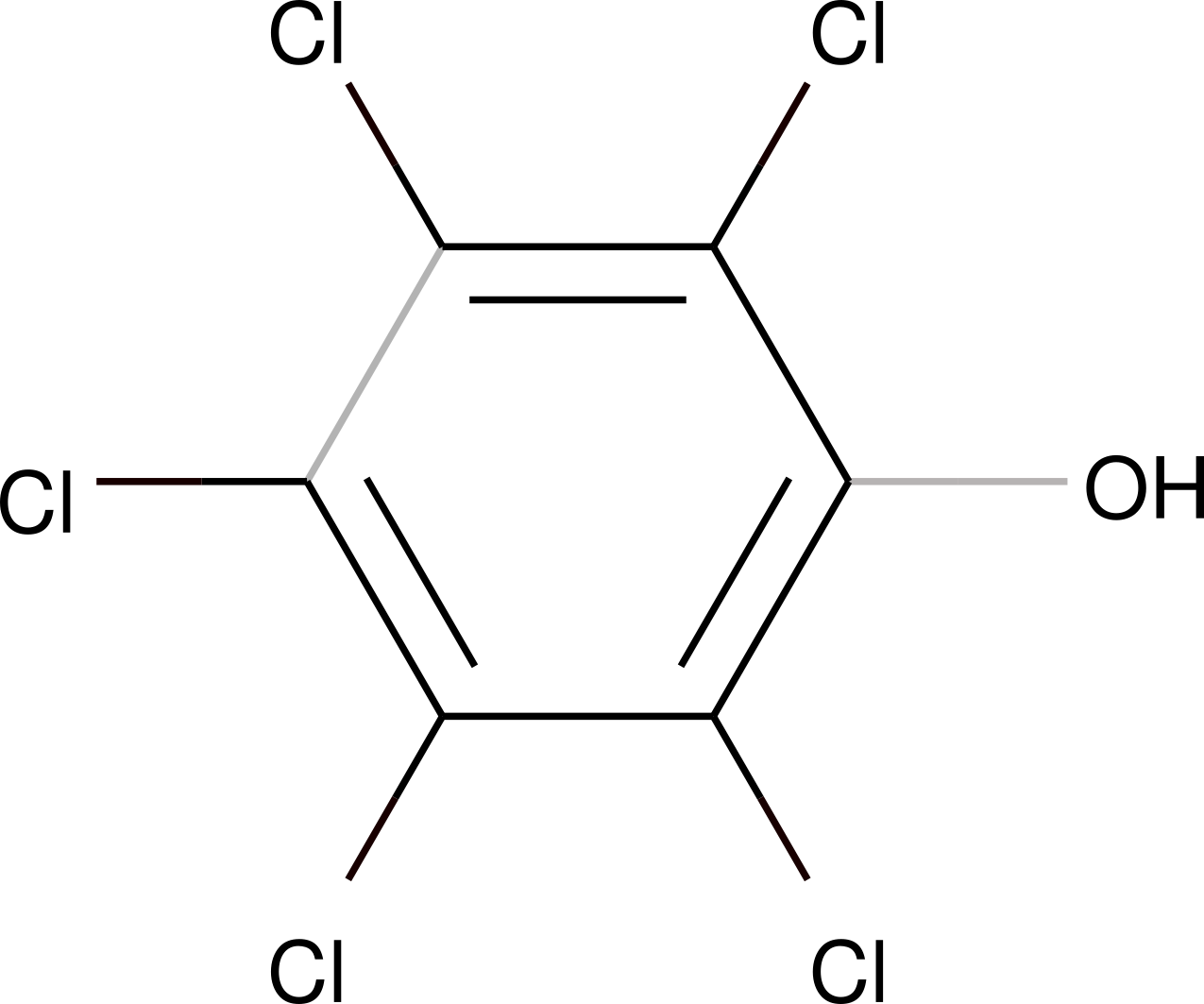}
    }
    \hspace{7pt}
    \subfigure[B1]{
         \centering
         \includegraphics[width=.35\linewidth]{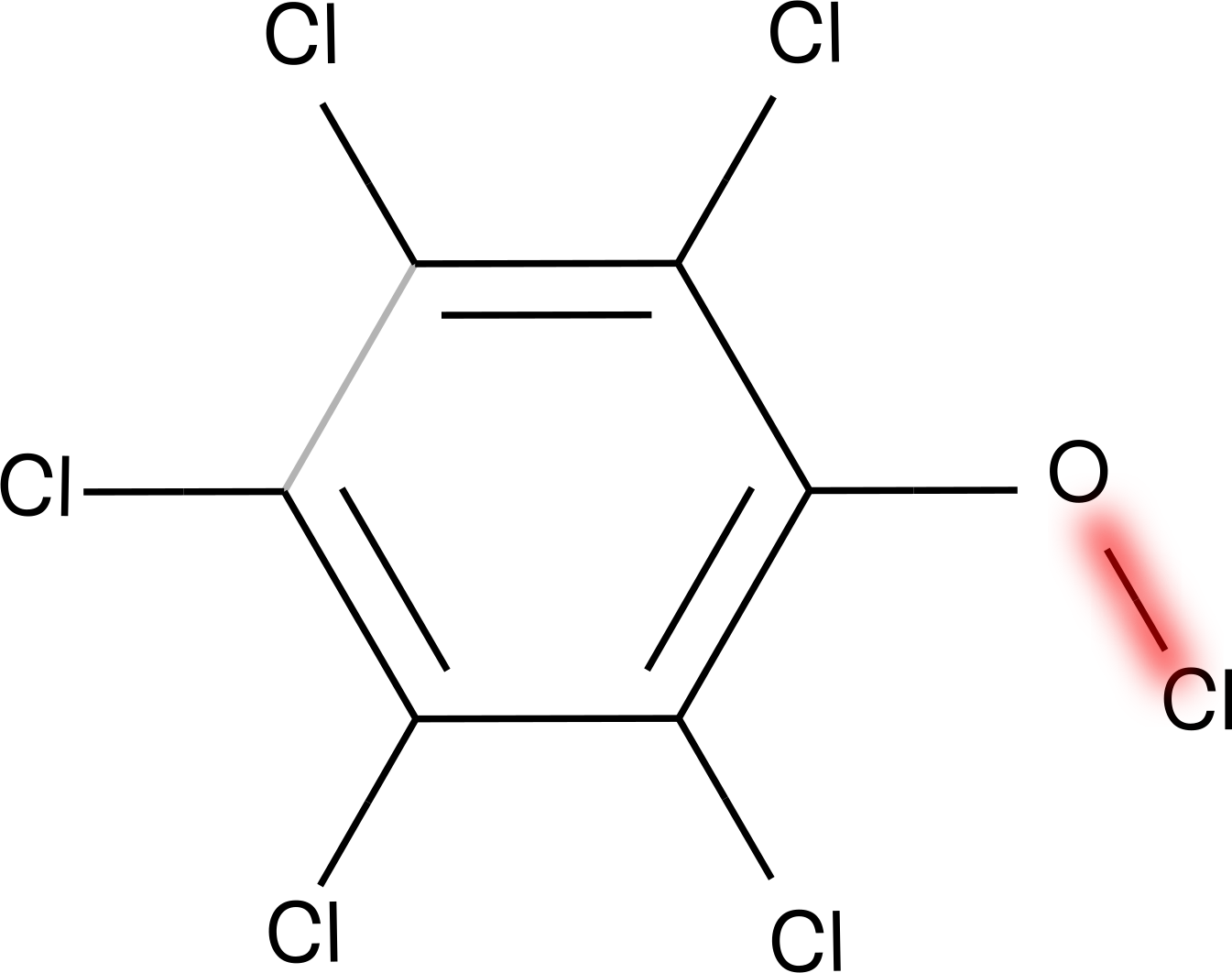}
    }
    \\[.1\baselineskip]
    \subfigure[B2]{
         \centering
         \includegraphics[width=.27\linewidth]{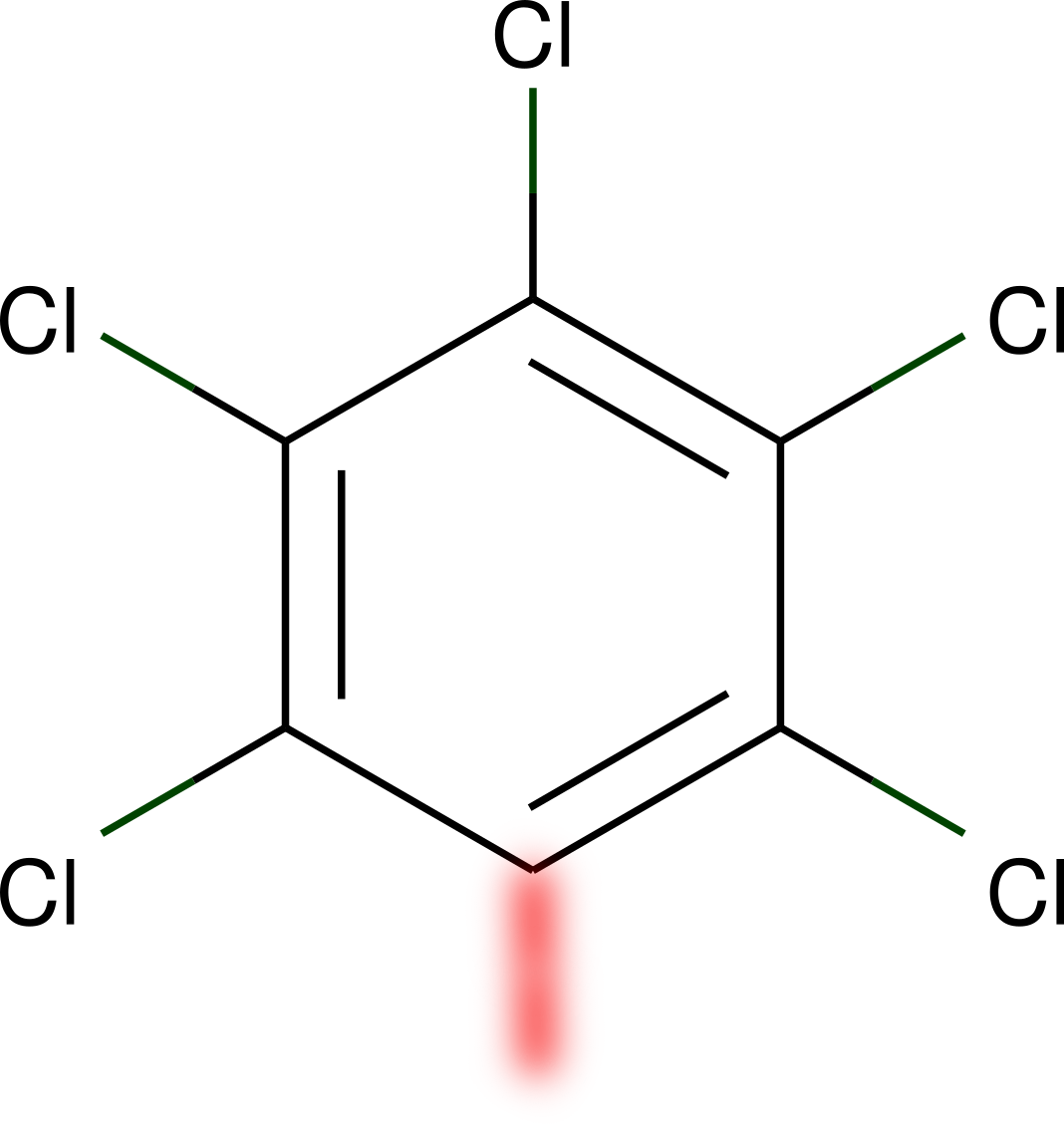}
    }
    \hspace{7pt}
    \subfigure[B3]{
         \centering
         \includegraphics[width=.35\linewidth]{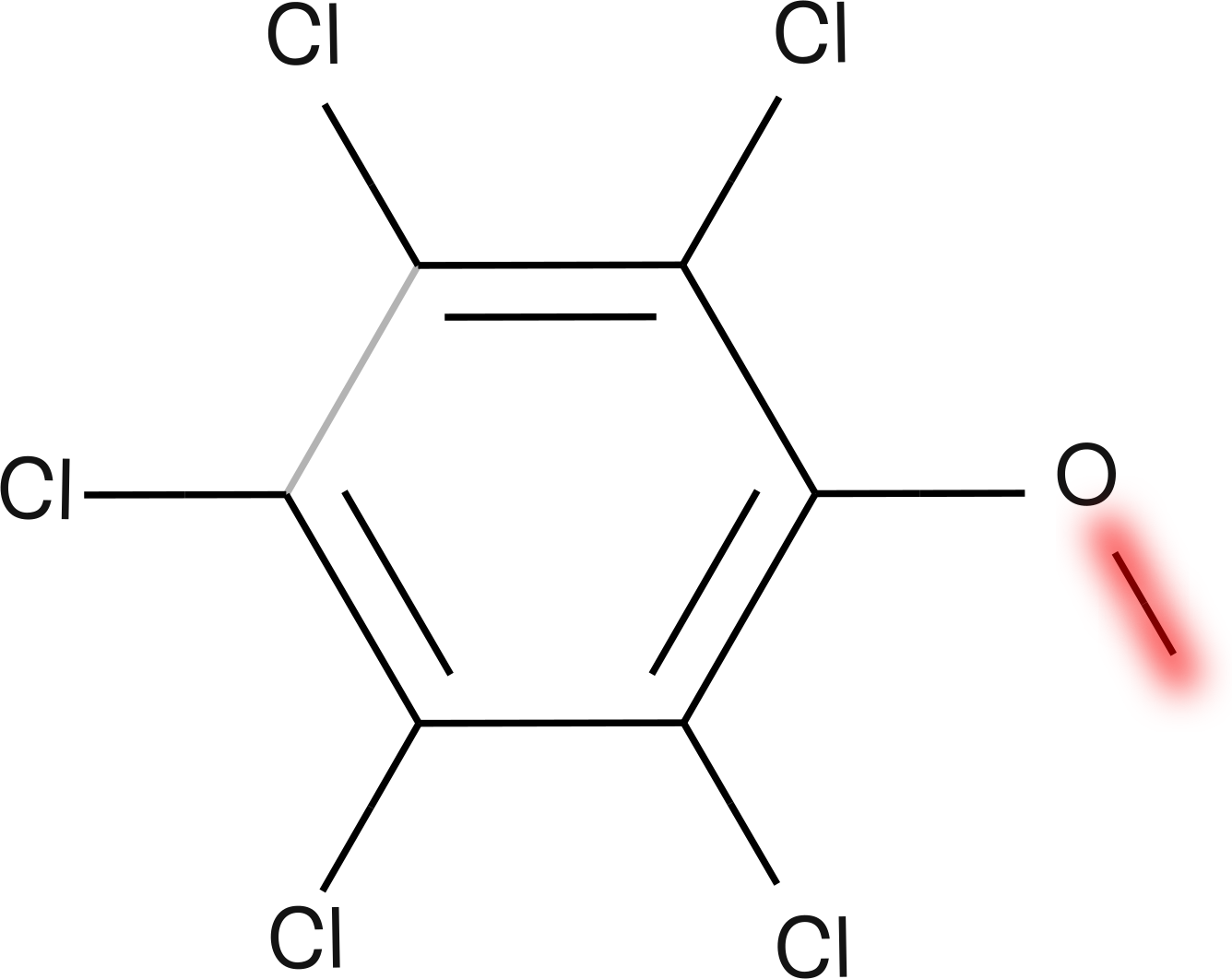}
    }
     \caption{ESOL sample alongside its counterfactuals (B1-3). Quantitative results are reported in \autoref{tab:res}.}
    \label{fig:ex2}
\end{wrapfigure}
We discuss a preliminary assessment of our explanations on two popular molecular property prediction benchmarks: Tox21 \cite{dortmund2016}, addressing toxicity prediction as a binary classification task, and 
ESOL \cite{wu2017moleculenet}, that is a regressive task on water solubility of chemical compounds.
Preliminarily, we scanned both datasets to filter non-valid chemical compounds. We considered structures to be valid molecules if they pass the RDKit \cite{landrum2006rdkit} sanitization check. In the end, Tox21 comprises 1900 samples, equally distributed among the two classes, while ESOL includes 1129 compounds.

The trained DGN comprises three
GraphConv \cite{morris2018weisfeiler} layers
with ReLu activations, whose hidden size is $256$
per layer for Tox21, and $128$ for ESOL. The network builds a layer-wise molecular representation via concatenation of max and mean pooling operations, over the set of node representations. 
The final neural encoding of the molecule is obtained by sum-pooling of the intermediate representations. This neural encoding is then feed to a three-layer feed-forward network, with hidden sizes of [128, 64, 32], to perform the final property prediction step. The trained DGNs achieved 87\% of accuracy and 0.52 MSE over the Tox21 and ESOL test sets, respectively. All experiments have been performed by using the Adam
optimiser with a learning rate of $5 \cdot 10^{-4}$.
During generation, we employed MEG to find the $5$ best counterfactual explanations for each molecule in test, ranked according to the multi-objective score in Section \ref{sect:model}. Ideally, we would like to observe counterfactual molecules that are structurally similar to the original compound while leading to a substantially different prediction. Due to the stringent page constraints, in the following we report two example explanation cases (one for each dataset). Further examples and results are available in the appendix.

\begin{wraptable}{l}{.6\textwidth}
    \vspace{-12pt}
    \footnotesize
    \centering \begin{tabular}{c c c c c c } Molecule & Target &
    Prediction & Similarity & Reward\\ \hline \\
    A0: \autoref{fig:ex1}
    & \texttt{NoTox} & {\tt NoTox} (0.70) & - & -\\
    A1: \autoref{fig:ex1}
    & - & {\tt Tox} (0.90) & 0.76 & 0.80\\
    A2: \autoref{fig:ex1}
    & - & {\tt Tox} (0.83) & 0.79 & 0.72 \\
    A3: \autoref{fig:ex1}
    & - & {\tt Tox} (0.80) & 0.68 &
    0.66\\ 
    B0: \autoref{fig:ex2}
    & -4.28 & -4.01 & - & -\\
    B1: \autoref{fig:ex2}
    & - & -6.11 & 0.29 & 1.14\\
    B2: \autoref{fig:ex2}
    & - & -5.93 & 0.31 & 1.11 \\
    B3: \autoref{fig:ex2}
    & - & -5.07 & 0.28 & 0.66 \\ 
    \end{tabular} \caption{
    Summary of preliminary results. A0 and B0 refers to molecules
    belonging to Tox21 and ESOL, respectively. Subsequent indexes
    refers to the related counterfactuals explanations.
    }  \label{tab:res}
\end{wraptable}
We present some quantitative result in \autoref{tab:res},
listing the best three counterfactual explanations collected, for both
tasks. We tested two similarity metrics: cosine similarity over the neural encodings, for Tox21, and the Tanimoto, for ESOL. Qualitative results are shown in \autoref{fig:ex1}
and \autoref{fig:ex2}. To ease the interpretation of our results, counterfactual modifications have been highlighted in red, while blurred edges represent those edges that have been masked out by GNNExplainer predictions. In other words, GNNExplainer interpretations are the sub-graphs formed by non-blurred edges.
As for the Tox21 sample, we evaluate MEG against a test molecule (i.e, A0) that has been correctly classified by the DGN as being non-toxic, outputting the counterfactuals A1-3 (i.e. molecules which the model considers toxic). We can see that the addition of a carbon atom may alter the DGN prediction, as shown by A1 and A2.
In fact, while A0 is classified correctly with 70\% certainty, A1-2
are predicted as toxic, with certainty of 90\% and 83\%, respectively.
Differently, A3 breaks the left side ring and achieves the lowest
neural encoding similarity score among the three, giving clues about potential substructure-awareness.
Furthermore, in \autoref{fig:ex1} we show how counterfactuals may help to detect inconsistencies in GNNExplainer predictions.
In fact, although GNNExplainer identifies the substructure CC(N)O as explanation for the original sample A0, MEG counterfactuals prioritize changes to different molecule fragments. These inconsistencies suggest that the GNNExplainer interpretation is too much targeted to the input molecule (A0) and does not generalize even for minor modifications of the input graph.

We now turn our attention to ESOL results (B0-3) shown in \autoref{tab:res}. B0 is an organic compound named pentachlorophenol, commonly used as a pesticide or a disinfectant, and is characterized by
nearly absolute insolubility in water. While the DGN achieved good predictive performance for its aqueous solubility value, the counterfactuals underlined that the $\varphi$-model predicted solubility decreases in case the oxygen atom is removed (e.g, B2), or modified somehow (e.g, B1, B3), highlighting how it is
highly relevant for the DGN prediction. As in the Tox21 sample, such relation is not adequately captured by GNNExplainer explanation for B0.
It is our hope that, based on our interpretability approach, an expert of the molecular domain could be able to gain a better insight into the whether the properties and patterns captured by the predictive model are meaningful from a chemical standpoint. 


\section{Conclusions}
We have presented MEG, a novel interpretability framework that tackles explainability in the chemical domain by generation of molecular counterfactual explanations. MEG can work with any DGN model as we only exploit input-output properties of such models. As a general comment of the preliminary results, one can note that while a local approach such as GNNExplainer may give good approximations when it comes to explaining the specific prediction, it lacks sufficient breadth to characterize the model behaviour already in a near vicinity of the sample under consideration. On the other hand, our counterfactual interpretation approach may find new samples which are likely to highlight the causes of a given model prediction, providing a better approximation to a locally interpretable model, e.g. B1-3 in \autoref{fig:ex1}. 
In conclusion, apart for its value in generating explanations that are well understood by a domain expert, MEG proposes itself both as a sanity checker for other local model explainers, as well as a sampling method to strengthen the coverage and validity of local interpretable explanations, such as in the original LIME method for vectorial data \cite{lime}.


%% file: appendix.tex
\begin{appendices}
\section{Additional Results}
\begin{wraptable}{r}{.6\textwidth}
    \vspace{-25pt}
    \footnotesize
    \centering \begin{tabular}{c c c c c c} Molecule & Target &
    Prediction & Similarity & Reward\\ \hline \\
    C0: \autoref{fig:ex3} 
    & -4.755 & -4.5195 & - & 1.57\\
    C1: \autoref{fig:ex3}
    & - & -2.6488 & 0.39 & 1.33 \\
    C2: \autoref{fig:ex3} 
    & - & -3.0170 & 0.65 & 1.12 \\
    
    D0: \autoref{fig:ex4}
    & {\tt NoTox} & {\tt NoTox} (0.71) & - & -\\
    D1: \autoref{fig:ex4}
    & - & {\tt Tox} (0.86) & 0.90 & 0.78 \\
    D2: \autoref{fig:ex4}
    & - & {\tt Tox} (0.80) & 0.91 & 0.73 \\
    
    E0: \autoref{fig:ex5}
    & {\tt Tox} & {\tt Tox} (0.78) & - & -\\
    E1: \autoref{fig:ex5}
    & - & {\tt NoTox} (0.94) & 0.69 & 0.86 \\
    E2: \autoref{fig:ex5}
    & - & {\tt NoTox} (0.84) & 0.89 & 0.73 \\
    \end{tabular}
    \caption{Summary of other preliminary results.}
    \label{tab:res2}
\end{wraptable}

\begin{wrapfigure}{l}{0.6\textwidth}
\centering
    \subfigure[E0]{
         \centering
         \includegraphics[width=.45\linewidth]{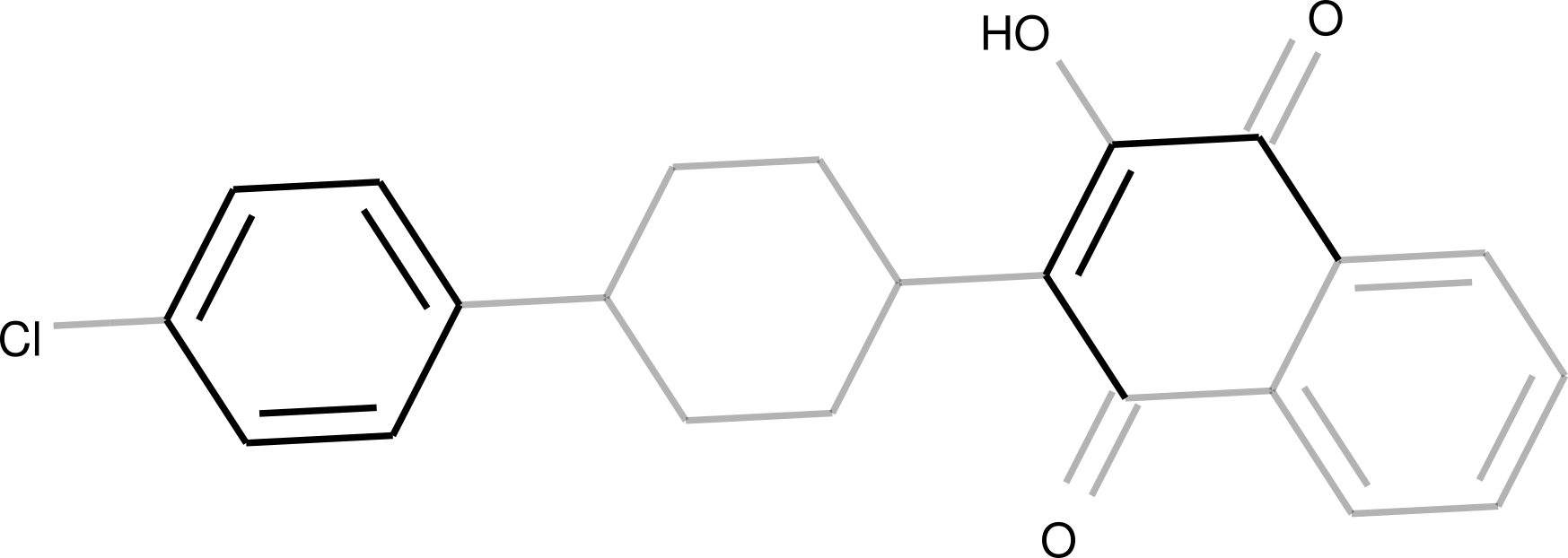}
    }\\
    \subfigure[E1]{
         \centering
         \includegraphics[width=.45\linewidth]{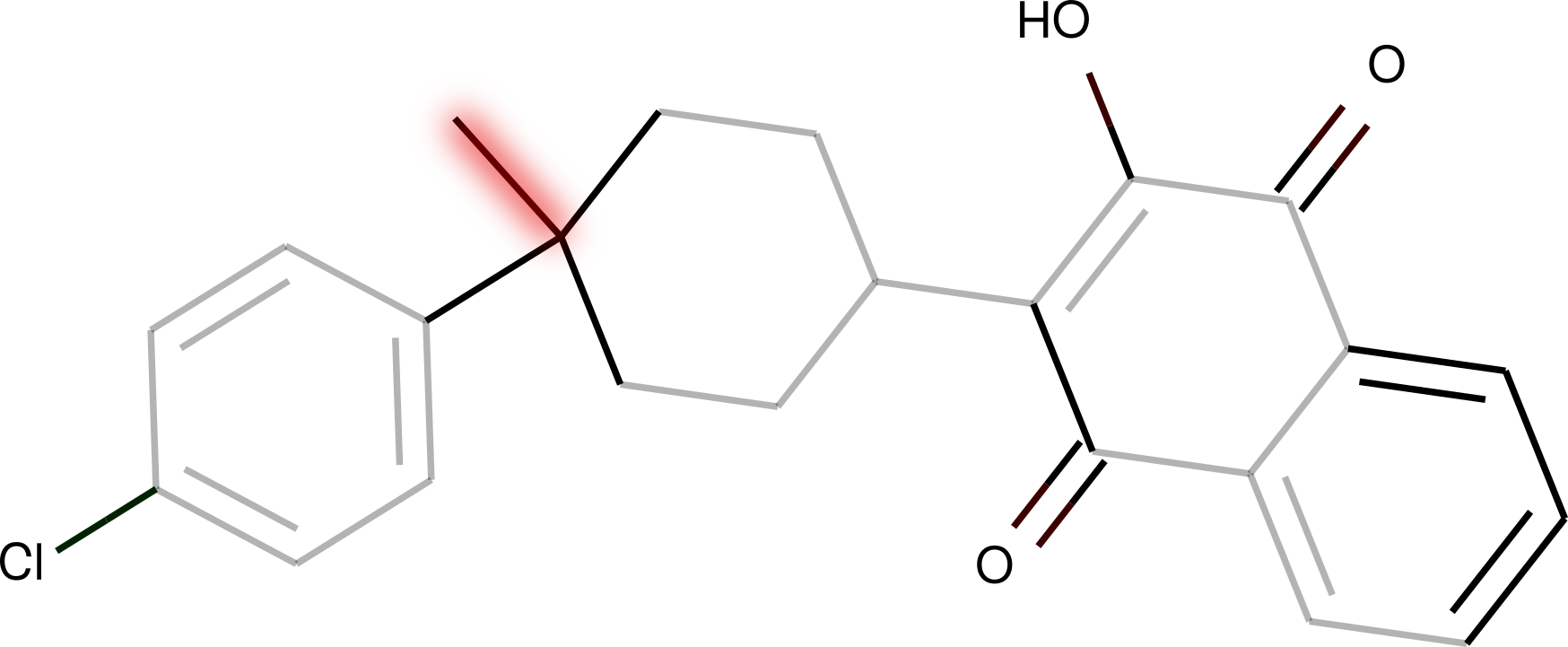}
    }
    \subfigure[E2]{
         \centering
         \includegraphics[width=.45\linewidth]{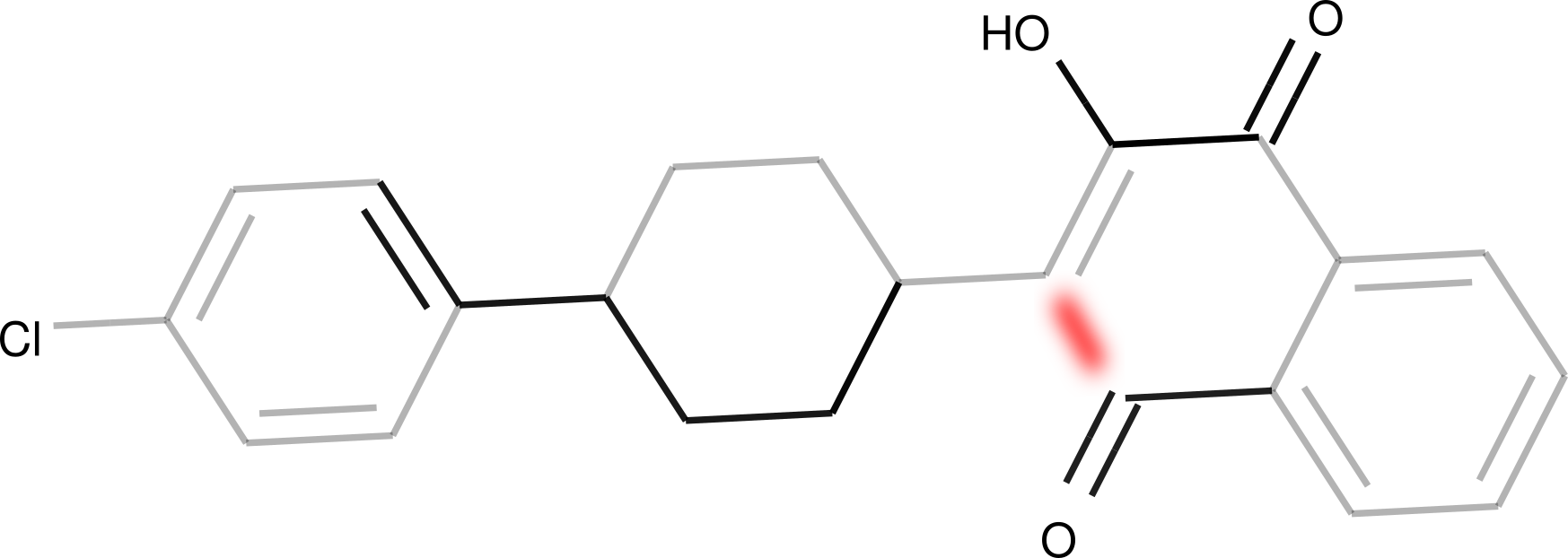}
    }
     \caption{E1 modifies the cyclohexane ring, which was not considered important in the explanation provided by GNNExplainer for the original molecule E0. E2 breaks the bond highlighted in red.}
    \label{fig:ex5}
\end{wrapfigure}

\begin{figure}[h]
    \begin{minipage}{.45\textwidth}
    \centering
    \begin{subfigure}
    \centering
    \subfigure[C0]{
         \centering
         \includegraphics[width=.45\linewidth]{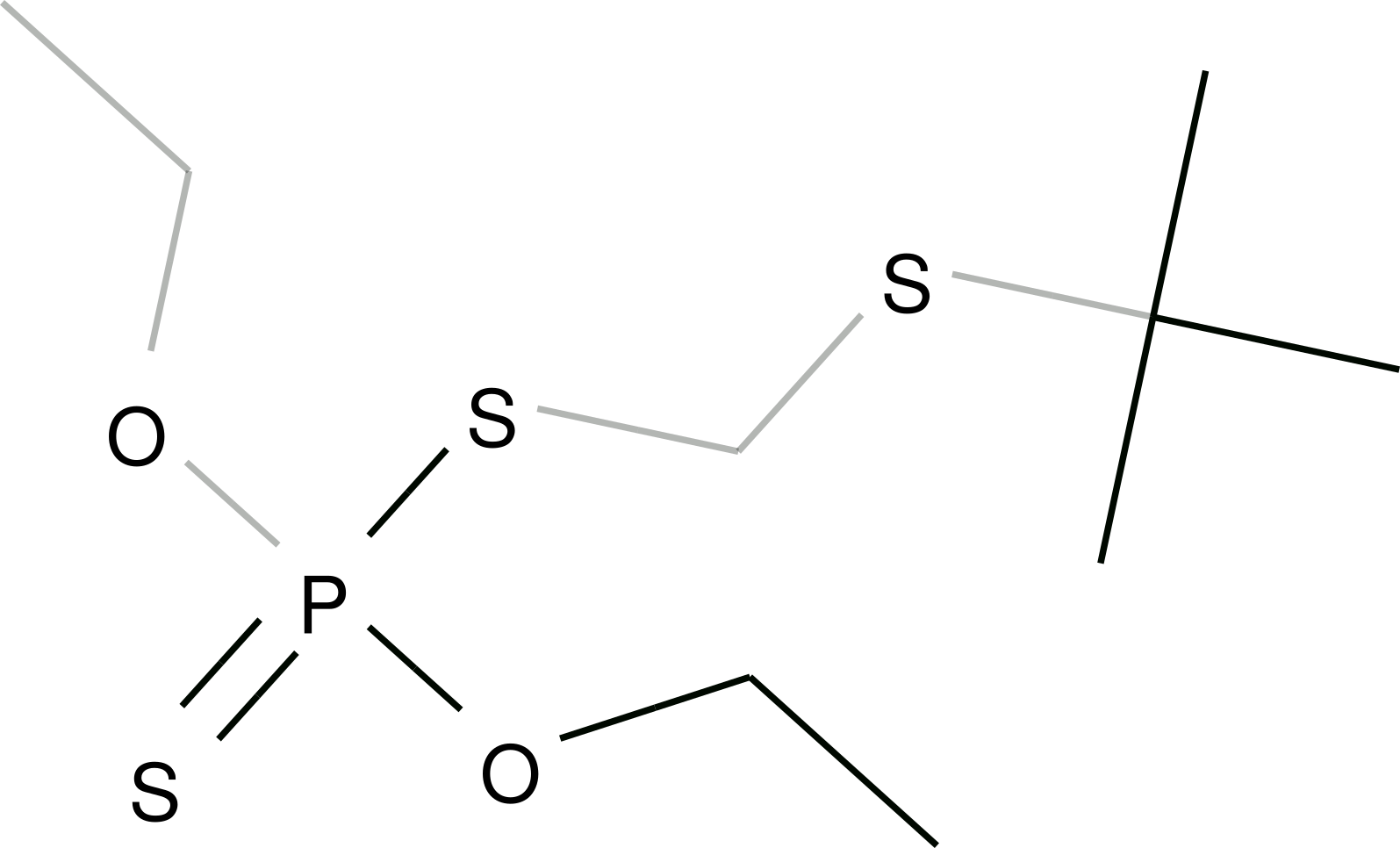}
    }\\[1.2\baselineskip]
    \subfigure[C1]{
         \centering
         \includegraphics[width=.45\linewidth]{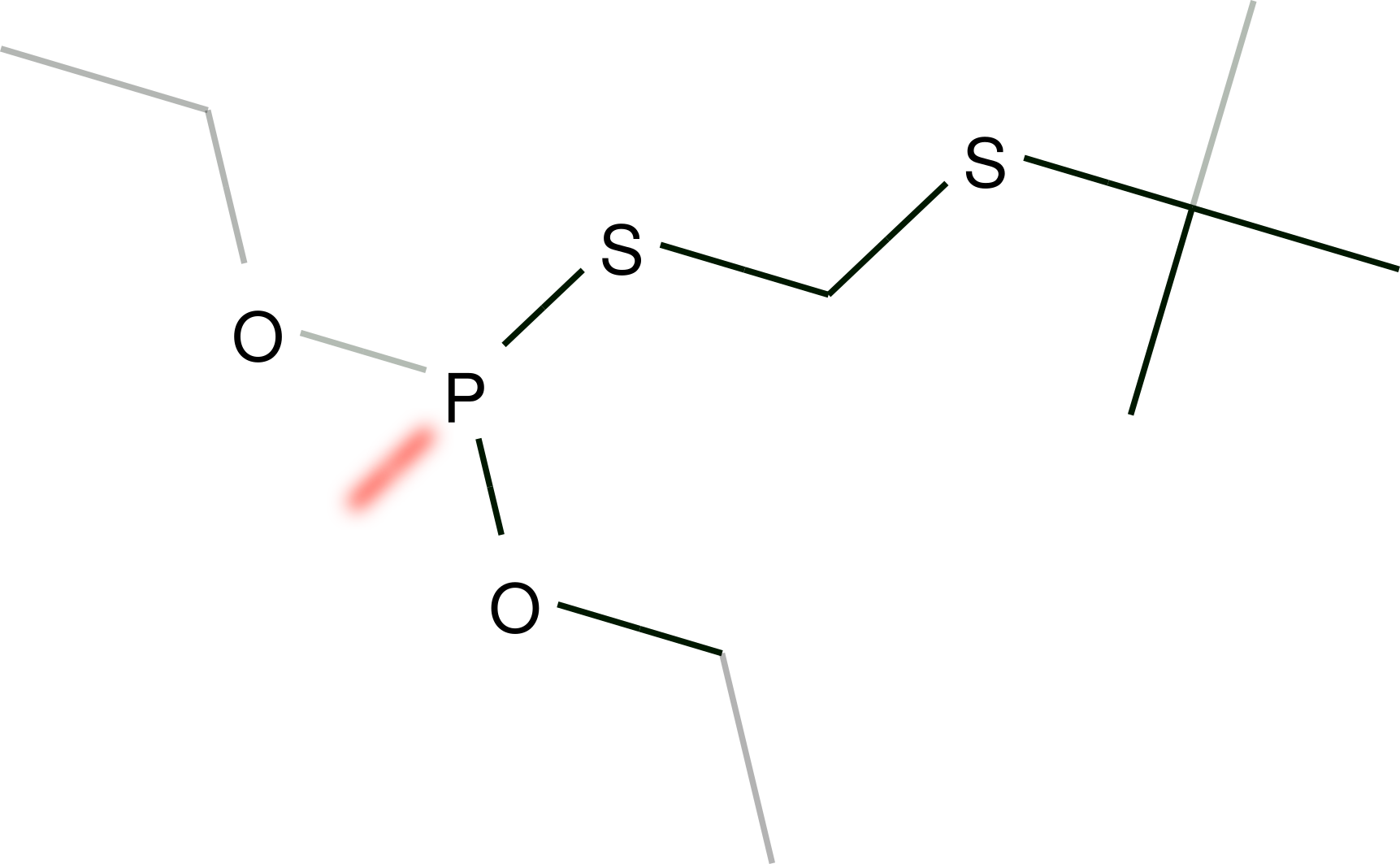}
    }
    \subfigure[C2]{
         \centering
         \includegraphics[width=.45\linewidth]{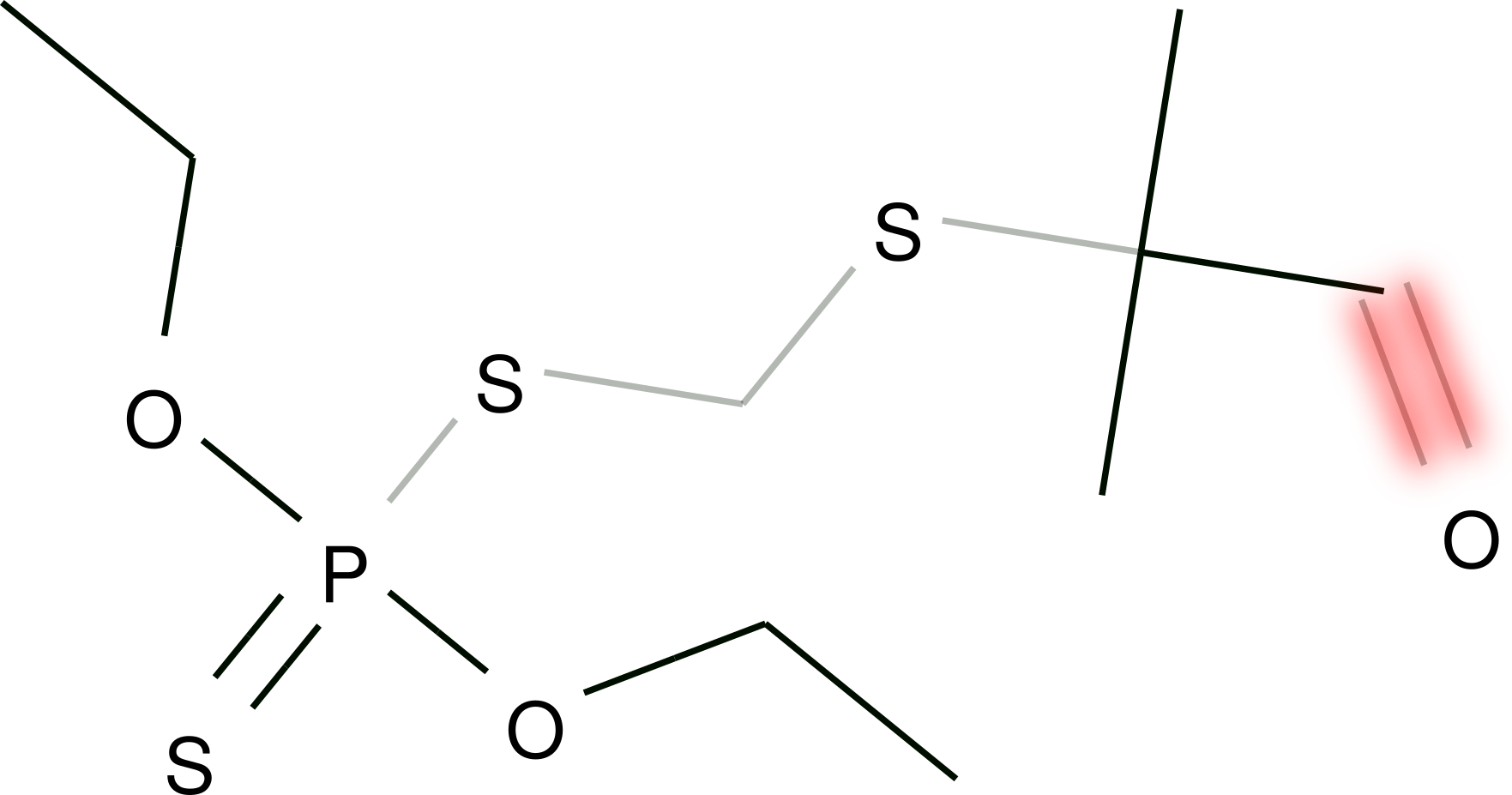}
    }
     \caption{ESOL. C1 removes the atom of sulphur. C2, instead, adds a new atom of oxygen and connect it to the molecule through a double bond.}
    \label{fig:ex3}
    \end{subfigure}
    \end{minipage}
    \hfill
    \begin{minipage}{.45\textwidth}
    \centering
    \vspace{-10pt}
    \begin{subfigure}
        \centering
        \subfigure[D0]{
         \centering
         \includegraphics[width=.45\linewidth]{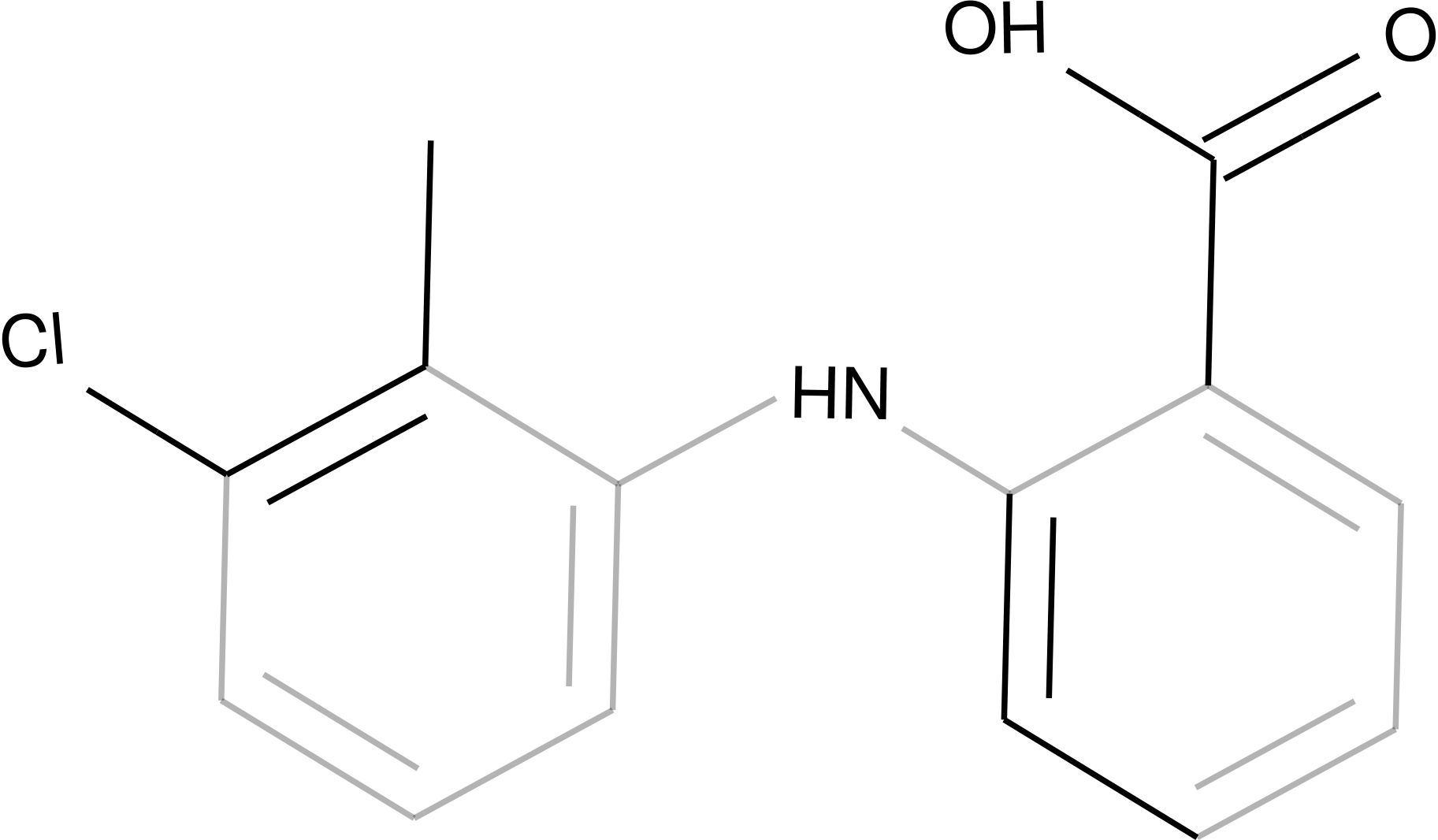}
    }\\
    \subfigure[D1]{
         \centering
         \includegraphics[width=.44\linewidth]{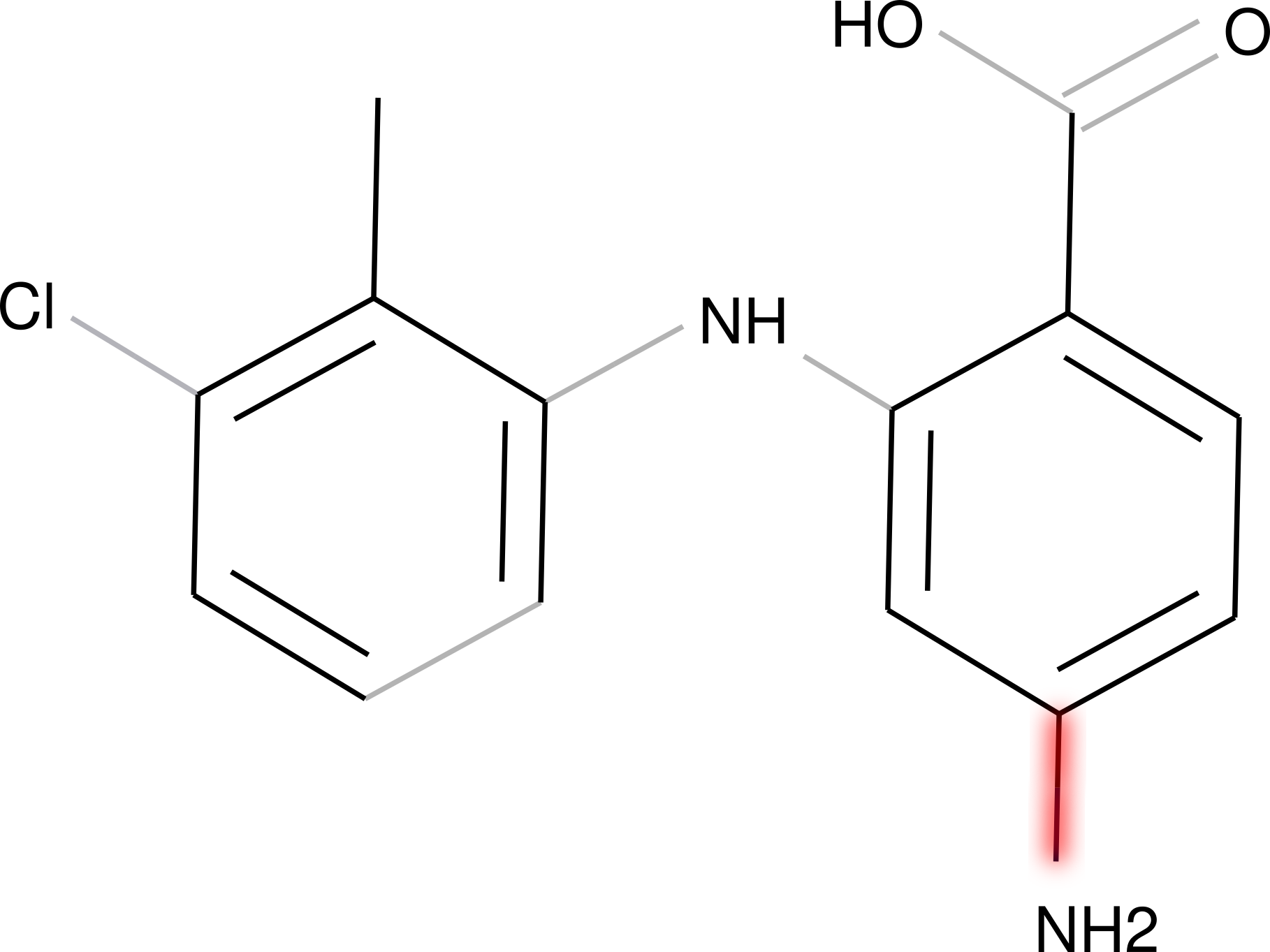}
    }
    \subfigure[D2]{
         \centering
         \includegraphics[width=.45\linewidth]{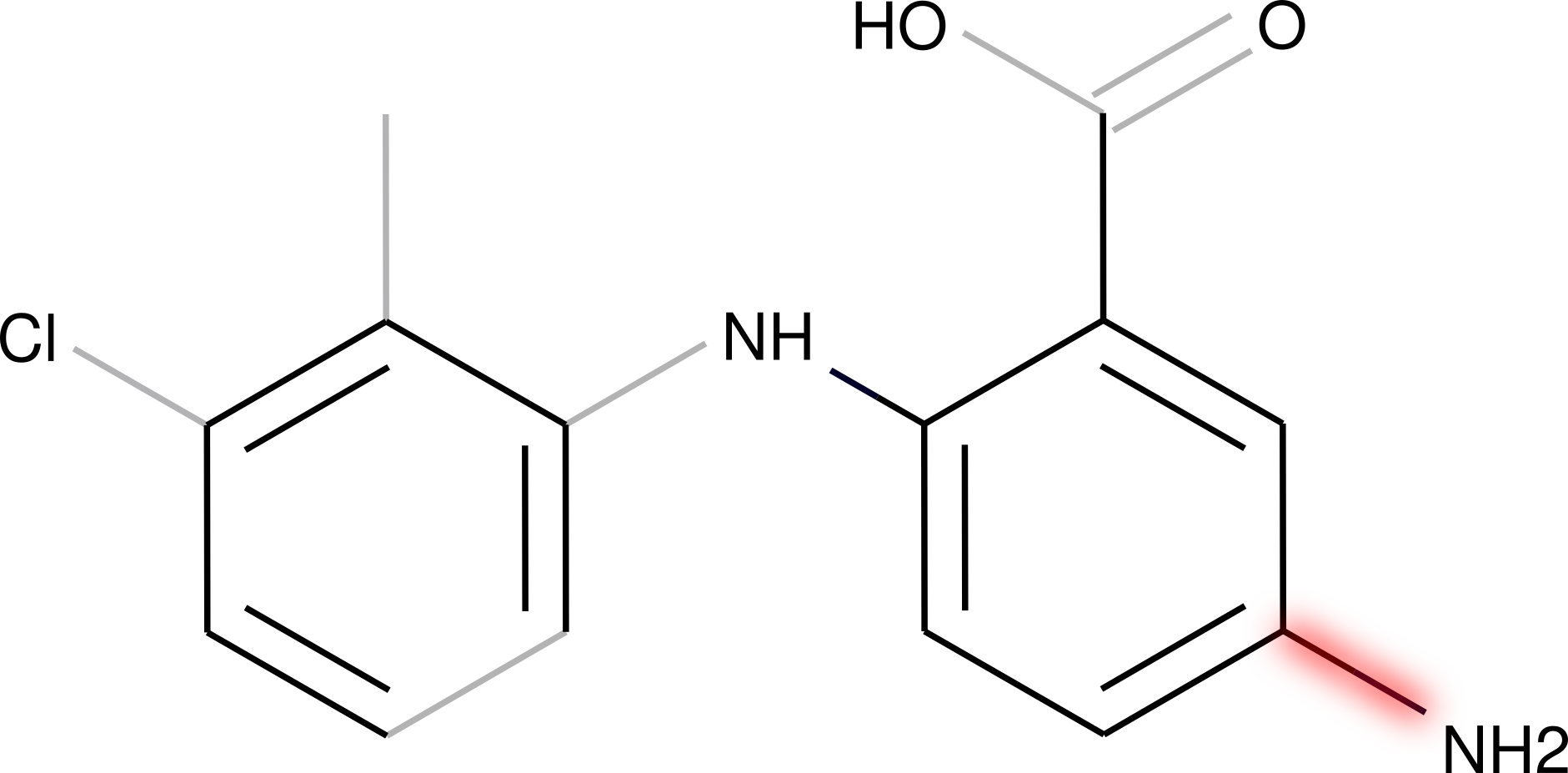}
    }
     \caption{Tox21. The agent adds atoms of nitrogen to the rightmost ring.}
    \label{fig:ex4}
    \end{subfigure}
    \end{minipage}
\end{figure}

\autoref{tab:res} provides experimental results for
three compounds, one of which belongs to ESOL (C0-2) and two to Tox21
(D0-2, E0-2). Visual feedback is shown in Figure \ref{fig:ex5}-\ref{fig:ex3}-\ref{fig:ex4}. 
As before, sharpness of graph edges indicates GNNExplainer explanations, while counterfactual modifications have been colored in red.

We seek for counterfactuals for an ESOL test compound, whose predicted solubility is close to the actual target. In this case, the atom of sulphur seems to have a negative impact on the predicted aqueous solubility. In this regard, C1 increases the compund solubility by removing, indeed, the atom of sulphur. In nature, a molecule of sulphur (i.e, S8 in SMILES encoding) is known to be insoluble. Such an analysis can drop preliminary hints about how the trained model may have learned such characteristics. 
Similarly to C1, C2 added an atom of oxygen causing the predicted water solubility to increase.

Another significant examples comprises D0-2. In fact, 
D0 is correctly classified as a non-toxic compound. However,
a simple addition of nitrogen makes the prediction change completely, resulting in classifying D1 and D2 as toxic with certainty of 86\% and 80\% respectively. Furthermore,
sanity checks on GNNExplainer explanation for D0 emphasize that D2 updates a blurred explanation fragment (i.e, the atom of carbon attached to the atom of nitrogen nor its incident bonds have been considered important in D0).
More interestingly, E0-2 present a potentially dangerous situation. In detail, starting from a toxic compound (E0),
E1 achieves to be recognized as non-toxic by simply adding an atom of carbon, and so does E2 by breaking one of the 
rings, as shown in \autoref{fig:ex5}.
In this case, the usefulness of our counterfactuals can be exploited to the fullest, highlighting such difficulties of the model under consideration which is crucial in real-world applications.

\end{appendices}